\documentclass{icrc29}
\usepackage{graphicx,amssymb,amsmath,times}
\begin{document}
\title[The Cake experiment  ...]{Cosmic ray abundance measurements with the CAKE balloon experiment}
\author[T. Chiarusi et al.] {S. Cecchini$^{1,4}$, T. Chiarusi$^2$, G. Giacomelli$^1$, S. Manzoor,$^{1,3}$, E. Medinaceli$^1$,
\newauthor
 L. Patrizii$^1$, V. Togo$^1$ \\
(1) Dip. Fisica dell'Universit\`a di Bologna and INFN, 40127 Bologna, Italy\\
(2) Dip. Fisica dell'Universit\`a "La Sapienza" di Roma, 00185 Roma, Italy\\
(3) PRD, PINSTECH, P.O. Nilore, Islamabad, Pakistan  \\
(4) INAF/IASF Sez. di Bologna, 40129 Bologna, Italy  \\
}
\presenter{Presenter: T. Chiarusi (tommaso.chiarusi@roma1.infn.it),   
ita-chiarusi-T-abs1-og11-oral }

\maketitle
\begin{abstract}
We present the results from the CAKE (Cosmic Abundance below Knee Energy) balloon
experiment which uses nuclear track detectors. The final experiment goal
is the determination of the charge spectrum of CR nuclei with Z $>$ 30  in the primary cosmic radiation.
The detector, which has a geometric acceptance of $\sim$ 1.7 m$^2$sr, was exposed in a trans-mediterranean stratospheric balloon flight. Calibrations of the detectors used (CR39 and Lexan), scanning strategies and algorithms for tracking particles in an automatic mode are presented. The present status of the results is discussed
\end{abstract}
\section{Introduction}
Precise measurements of primary cosmic ray (CR) nuclei abundances
beyond Z$>$30 could allow to discriminate between two competing models
\cite{b@fip} \cite{b@volatile} proposed for the composition at their sources. Especially ratios among heaviest elements appear to be the best discriminant for most current theories regardless of the propagation history of these particles \cite{b@wadd}. 
In order to be able to measure these ratios, high charge resolution and high statistics are needed since the flux of heavy nuclei is very low ($<$ 1 part/m$^2$h).
Even if most measurements of the primary high energy cosmic ray
composition are indirect \cite{b@eastop}, untill now, 
the largest statistics above the Fe threshold came from
direct experiments of the  HEAO-C3/Ariel-6 mission \cite{b@heao} \cite{b@ariel},
which unfortunately suffered from low charge resolution.  
Recently, the TIGER balloon experiment \cite{b@tiger} published new results in the range of 30 $<$ Z $<$ 40, after a flight of 31 days over Antarctica.
The use of plastic nuclear track detectors (NTDs) onboard space
vehicules to study the radiation environment is well known
\cite{b@LDEF}\cite{b@Trek}. The use of NTDs makes possible to build
experiments with large geometrical factors, and balloon flights of
20-40 days appear feasible from Antartica or in north circumpolar
routes. These allow to study the Z range 30-40.
We have proposed to the Italian Space Agency (ASI) an R\&D programme with the final goal of using a large experiment (typically 8m$^2$sr) in a long duration balloon flight. CAKE is a prototype experiment, fully described in \cite{b@ceccocake}, which uses stacks of CR39 and Lexan passive nuclear track detectors exposed in a short balloon flight. It was flown for the first time in 1999 and preliminary results  (in the range 5$<$Z$<$30) have been presented in \cite{b@icrc03}\cite{b@barca}. 
In this paper we report on new approaches used for automatic scanning and new analys methods of the recorded events. In particular we describe  new algorithms implemented in the tracking and selection software. Preliminary results obtained with this technique are also presented. 
 
\section{The Test Flights}
A first test flight was carried out in July 1999. The balloon was
launched from the Trapani-Milo base (12$^\circ$.50E, 32$^\circ$.92N)
of ASI and landed in central Spain after 22 hours. The plafond
altitude was 37-40 km (3.5-3 g cm$^{-2}$) for 17.8 hours. Along the
trajectory the average vertical rigidity cut-off was about 8 GV
(introducing a cut off for the minimal energy of the impinging CRs of
about 3 GeV/nucleon ) \cite{b@SS}. The gondola was not azimuthally controlled. The air temperature inside the cylinders that contained the NTD stacks was never higher than 34$^\circ$ C.
A second flight was attempted from the same base in 2001 but the balloon was unable to reach the plafond after 4 hours, so the flight was interrupted.
\begin{figure}[h]
\begin{center}
\includegraphics*[width=1.\textwidth,angle=0,clip]{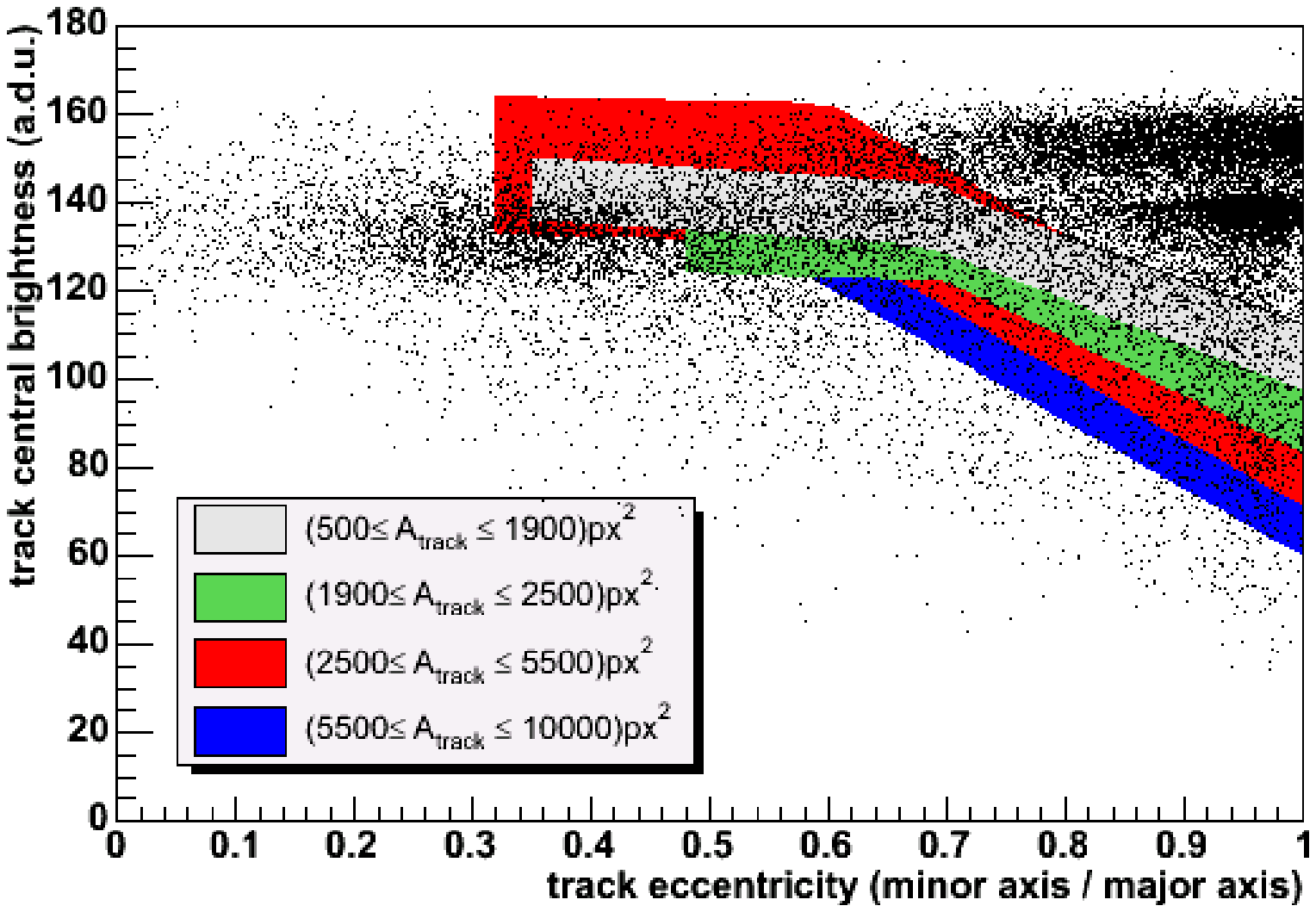}
\caption{\label {fig1} The off-line pre-selection fiducial belts (the
  coloured stripes) in the brightness versus eccentricity plot, with
  the measured raw data superimposed. It is apparent that the bulk of
  events with  eccentricity of aboout 1  are purely background.}
\end{center}
\end{figure}

\section{The Calibrations} 
The calibrations used for the CR39 stacks are the ones described in \cite{b@Giac}. In order to improve the contrast when using automatic scan a new set of calibrations were performed using 1 A GeV $^{26+}$Fe from the BNL AGS and  158 A GeV $^{49+}$In beams at the CERN SPS. Lexan detectors have been calibrated using 30 A GeV $^{82+}$Pb beams at the CERN SPS. We are also experiencing new etching conditions in order to avoid the formation of too many background tracks (low Z or end-of-range nuclei) and plastic surface defects. 
\section{Scanning strategy and track reconstruction algorithms}
In order to distinguish heavy CR track candidates, which are very rare, from background in CR39 (which are sensitive to Z/$\beta$ $>$5) we need to develop an efficient data acquisition and data analysis strategy, possibly in automatic mode.
In our standard procedure several ($>$3) consecutive CR39 foils belonging to the same stack are etched in a 6N NaOH solution at 70$^\circ$ for 40 hours.
The plates are then scanned in automatic mode with the ELBEK Image Analyzer System \cite{b@elbek}. During this step the position and major and minor axes size are measured on the top surface and stored. 
An off-line pre-selection of the recorded raw data was applied to remove those tracks having brightness and eccentricity outside prefixed fiducial bands. These bands were determined from an independent set of "good" data, measured with interactive scan sessions at the microscope.
At present we are developing a Multilayer Perceptron neural network for implementing the off-line event filter. Furthermore, in order to keep the alignment among the plates (during the scanning operations), to within $\sim$30 $\mu$m, we implemented a new microscope stage. 
A dedicated tracking algorithm was applied in order to select tracks
crossing all the selected plates, since CR nuclei are expected to be highly penetrating.
Each candidate etch-pit was tracked individually, by means of a {\it fiducial area recursive method}.
\begin{figure}[h]
\begin{center}
\includegraphics*[width=1.\textwidth,angle=0,clip]{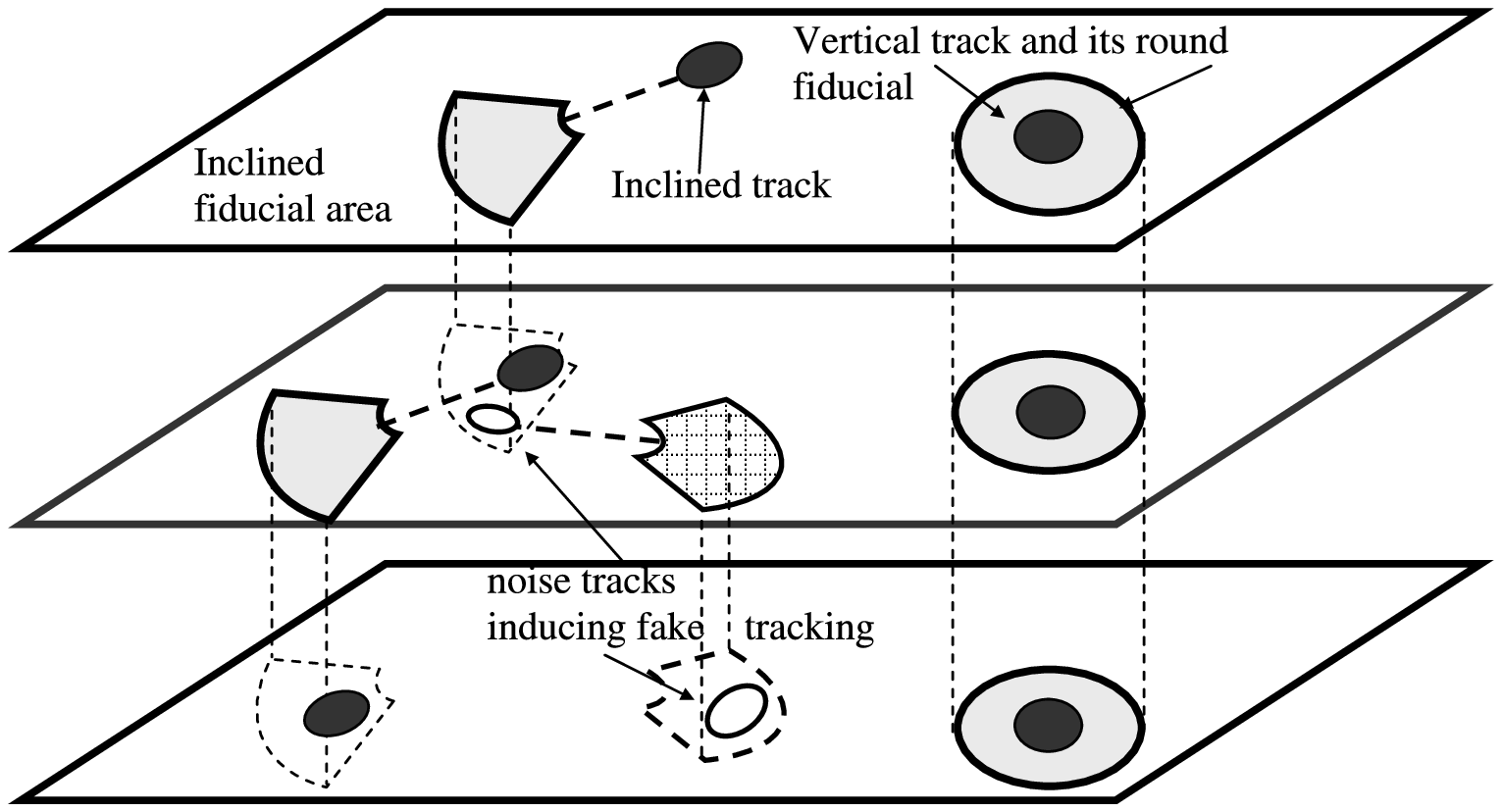}
\caption{\label {fig2}  A sketch of the  tracking procedure: according
  to the inclination and orientation of a first track on the uppermost
  foil, a fiducial area is  determined where to look for a counterpart
  track on the lower foils. On the left, the inclined track originates
  a sector  area, while, on  the right, a  circular area is set  for a
  vertical  track. The tracking  goes on  recursively through  all the
  scanned foils belonging to the same stack.  Also a noise track falls
  in one fiducial area (see left side of figure) when tracking an inclined track, introducing a
  fake alternative to the possible final trajectories. However, the wrong thread
will  be  removed with  the  post-selection  by  requiring a  linear
  trajectory for the impinging particles.}
\end{center}
\end{figure}
Starting from a track on the uppermost foil of the stack, and knowing the track inclination and orientation with respect to the plate, the algorithm defines a restricted area where to look for the counterpart track on the foil below. This procedure is made recursively through the foils of the stack. A full trajectory is determined when, for each scanned foil one track is linked with its upper and lower counterparts, yielding a {\it family} of tracks through the stack itself. The uncertainties on the size of each fiducial area arise from the uncertainties on the measured etch-pit base ($\pm$10 $\mu$m) on the thickness of each foil (measured with a maximum error of $\pm$ 5$\mu$m) and on their mutual alignement in the stack. We estimated a maximum shift of about 30 $\mu$m along the x-axis of the stage holding the NTD foils.
The first scan usually produces more than 10000 raw tracks per foil. After tracking through 3 plates the selected events are reduced to roughly 1000 candidates.

\section{Results and their comparison with simulations}
The results here reported come from a scanned area of $\sim$870 cm$^2$, representing $\sim$15\% of the total detector. 
 \begin{figure}[h]
\begin{center}
\includegraphics*[width=1.\textwidth,angle=0,clip]{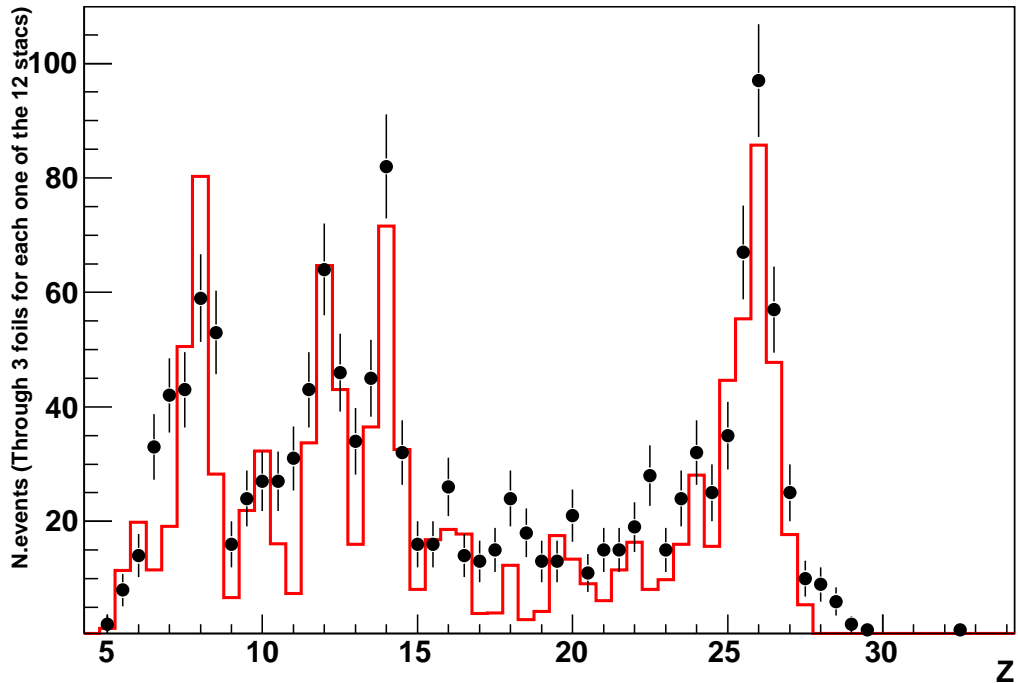}
\caption{\label {fig3} Charge spectrum of the 1409 events from the
  measurement of 15\% area of the CAKE detector. Data (black dots) with statistical errors are compared to the MonteCarlo simulations (histogram).}
\end{center}
\end{figure}

Fig. \ref{fig3} shows the charge spectrum in the range 5$\leq$Z$\leq$40 for events tracked through three consecutive plates compared to the MonteCarlo expectations  \cite{b@phd}. The total number of reconstructed events is 1409.
In the simulation we took into account the residual atmosphere and the detector's, scanning, tracking and selection efficiencies.

\section{Conclusions}
We are now in the process of completing the analysis of the remaining
sheets of the CAKE experiment. The scanning and tracking automatic
procedures that we have developed assure the possibility of making the
measurements with large area NTDs with good efficiency in a reasonable
time. 
From the present results (50\% more data measured after the ICRC conference agree with
what stated here) we are confident that, with a single
balloon flight of 30 day duration or longer of an 8 m$^2$sr detector
similar to CAKE, it will be possible to reach the sensitivity, charge
resolution and statistics, needed for discriminating between the
different scenarios of the origin of cosmic rays. It should also lead
to a good search for exotic heavy particles in the primary cosmic
radiation \cite{b@balestra}

\section{Acknowledgements}
We acknowledge the collaboration of E. Bottazzi, L. Degli Esposti, D. Di Ferdinando,
M. Errico, M. Frutti, G. Grandi and C. Valieri in the processing of NTDs and support with ELBEK. The Servizio Officina of  INFN, Sezione Bologna, helped in designing and building the CAKE experiment. We thank also S. Di Maria and D. Fiorini for their contribution to the off-line software development. We finally acknowledge the support of the Trapani-Milo Stratospheric Balloon Launch Base of ASI.  T. Chiarusi had been partially supported by an ASI fellowship.

\end{document}